\documentclass[ALICE,manyauthors]{cernphprep}
\usepackage[comma,square,numbers,sort&compress]{natbib}

\usepackage{lineno}
\usepackage{xspace}
\usepackage{hyperref}
\usepackage[T1]{fontenc}
\usepackage{orcidlink}
\newcommand{\ee}           {\mbox{$\rm e^{+}e^{-}$}\xspace}

\newcommand{\pt}           {\ensuremath{p_{\rm T}}\xspace}

\newcommand{\nineH}        {$\sqrt{s}~=~0.9$~Te\kern-.1emV\xspace}
\newcommand{\seven}        {$\sqrt{s}~=~7$~Te\kern-.1emV\xspace}
\newcommand{\twoH}         {$\sqrt{s}~=~0.2$~Te\kern-.1emV\xspace}
\newcommand{\twosevensix}  {$\sqrt{s}~=~2.76$~Te\kern-.1emV\xspace}
\newcommand{\five}         {$\sqrt{s}~=~5.02$~Te\kern-.1emV\xspace}
\newcommand{\twosevensixnn}{$\sqrt{s_{\mathrm{NN}}}~=~2.76$~Te\kern-.1emV\xspace}
\newcommand{\fivenn}       {$\sqrt{s_{\mathrm{NN}}}~=~5.02$~Te\kern-.1emV\xspace}

\newcommand{\GeVc}         {Ge\kern-.1emV$/c$\xspace}
\newcommand{\MeVc}         {Me\kern-.1emV/$c$\xspace}
\newcommand{\TeV}          {Te\kern-.1emV\xspace}
\newcommand{\GeV}          {Ge\kern-.1emV\xspace}
\newcommand{\MeV}          {Me\kern-.1emV\xspace}
\newcommand{\GeVmass}      {Ge\kern-.2emV/$c^2$\xspace}
\newcommand{\MeVmass}      {Me\kern-.2emV/$c^2$\xspace}

\newcommand{\Lambdac}{\Lambda_{\rm c}^{+}}
\newcommand{\Xicplus}{\Xi_{\rm c}^{+}}
\newcommand{\Xiczero}{\Xi_{\rm c}^{0}}
\newcommand{\Xiczeroplus}{\Xi_{\rm c}^{0,+}}
\newcommand{\Sigmac}{\Sigma_{\rm c}^{0,++}}
\newcommand{\Dzero}{\rm D^{0}}
\newcommand{\Ds}{\rm D^{+}_{s}}
\newcommand{\Omegac}{\Omega_{\rm c}^{0}}

\begin{document}

\begin{titlepage}
\PHyear{2022}       
\PHnumber{106}      
\PHdate{23 May}  

\title{First measurement of $\rm \Omega_c^0$ production in pp collisions at $\sqrt{s}=13$~TeV}
\ShortTitle{$\rm \Omega_c^0$ production in pp collisions at $\sqrt{s}=13$~TeV}   

\Collaboration{ALICE Collaboration\thanks{See Appendix~\ref{app:collab} for the list of collaboration members}}
\ShortAuthor{ALICE Collaboration} 

\begin{abstract}
The inclusive production of the charm--strange baryon $\Omegac$ is measured for the first time via its hadronic decay into $\rm \Omega^-\pi^+$ at midrapidity ($|y|<0.5$) in proton--proton (pp) collisions at the centre-of-mass energy $\sqrt{s}=13$~TeV with the ALICE detector at the LHC. The transverse momentum ($p_{\rm T}$) differential cross section multiplied by the branching ratio is presented in the interval $2<p_{\rm T}<12~{\rm GeV}/c$. The $p_{\rm T}$ dependence of the $\rm \Omega_c^0$-baryon production relative to the prompt $\rm D^0$-meson and to the prompt $\rm \Xi_c^0$-baryon production is compared to various models that take different hadronisation mechanisms into consideration. In the measured $\pt$ interval, the ratio of the $\pt$-integrated cross sections of $\rm \Omega_c^0$ and prompt $\Lambdac$ baryons multiplied by the $\rm \Omega^-\pi^+$ branching ratio is found to be larger by a factor of about 20 with a significance of about 4$\sigma$ when compared to $\rm e^+e^-$ collisions.

\end{abstract}
\end{titlepage}

\setcounter{page}{2} 


Recent measurements of charm-baryon production at midrapidity by the ALICE Collaboration~\cite{ALICE:2020wla, ALICE:2020wfu, ALICE:2021rzj, ALICE:2021psx, ALICE:2021bli} show that the $\Lambdac/\Dzero$, $\Xiczeroplus/\Dzero$, and $\Sigmac/\Dzero$ baryon-to-meson yield ratios are higher in pp collisions at LHC energies than in $\rm e^+e^-$ collisions, indicating that charm hadronisation occurs via different processes in the two collision systems~\cite{ALICE:2021dhb}. The ratios are found to decrease with increasing transverse momentum ($\pt$), a trend not expected by models based on factorisation and on the usage of the fragmentation functions extracted from \ee collisions. A significant dependence of the $\pt$-differential $\Lambdac/\Dzero$ ratio with the multiplicity of charged particles produced in the event was also observed in pp collisions at $\sqrt{s}=13$~TeV~\cite{ALICE:2021npz}, possibly suggesting a continuous evolution of this ratio from low-multiplicity pp collisions to the highest multiplicity of charged particles characterising Pb--Pb collisions with a small impact parameter~\cite{ALICE:2021bib}.

Higher charm baryon-to-meson ratios in pp collisions with respect to $\rm e^+e^-$ collisions are expected by models that either include dynamical processes that are relevant in quark-and-gluon enriched systems (e.g. colour reconnection beyond leading colour approximation~\cite{Christiansen:2015yqa} and quark coalescence~\cite{Minissale:2020bif}), or that treat hadronisation as a statistical process~\cite{Song:2018tpv,He:2019tik}. 

The Lund string fragmentation model~\cite{Andersson:1983jt,Andersson:1997xwk} implemented in the PYTHIA event generator~\cite{Sjostrand:2006za,Sjostrand:2007gs,Sjostrand:2014zea}, is one of the main hadronisation models used in general-purpose Monte Carlo event generators~\cite{Buckley:2011ms}. 
In the default version of PYTHIA~8 (Monash 2013 tune~\cite{Skands:2014pea}), the choice of quarks and gluons that are matched to form strings, encoding colour-confining potentials, is done in the leading-colour approximation. This configuration suppresses the connection of quarks and gluons coming from independent parton scatterings, realising heavy-quark fragmentation and hadronisation schemes very similar to those occurring in $\rm e^+e^-$ collisions. As a result, all of the baryon-to-meson ratios mentioned above are severely underestimated. The extension of colour reconnection beyond the leading colour (CR-BLC) approximation~\cite{Christiansen:2015yqa} allows the calculations to better approximate quantum chromodynamic colour algebra when matching partons to form strings and enhances the role of ``junction" colour-topologies that favour the formation of baryons. 
The CR-BLC model reproduces the $\Lambdac/\Dzero$ ratio, including the dependence on event multiplicity~\cite{ALICE:2021npz}, and the $\Sigmac/\Dzero$ ratio~\cite{ALICE:2021rzj}, but it underestimates the $\Xiczeroplus/\Dzero$~\cite{ALICE:2021psx, ALICE:2021bli}.

In the Catania model~\cite{Minissale:2020bif}, charm quarks can hadronise via ``vacuum"-like fragmentation as well as recombine (coalesce) with surrounding light quarks from the underlying event. The Wigner formalism is used to calculate the probability to form a baryon (meson) given the phase-space distribution of three (two) quarks. Within uncertainties, this model reproduces the charm baryon-to-meson ratios measured so far in pp collisions, though it tends to systematically underpredict the $\Xiczeroplus/\Dzero$ and the $\Xiczeroplus/\Sigmac$ ratios. 

In the models implementing hadronisation on a statistical basis, the relative abundances of the various charm-hadron species are determined by statistical weights that depend on the hadron mass, spin, and on the system properties. The $\pt$ dependence of the predicted ratios can have different origins. It derives from the feed-down from higher-mass state decays in the model of Ref.~\cite{He:2019tik}, in which a large set of not-yet-observed charm-baryon states is assumed, following the expectation of the relativistic quark model~\cite{Ebert:2011kk}. In the quark-recombination model (QCM)~\cite{Song:2018tpv} it instead derives from the requirement that charm quarks form hadrons by combining with light quarks with the same velocity. Both models describe the $\Lambdac/\Dzero$ and $\Sigmac/\Dzero$ ratios and underestimate the $\Xiczeroplus/\Dzero$ ratio in pp collisions, with the QCM prediction being closer to the data, although lower by about a factor of two.

The $\Omegac$ baryon is composed of a charm quark and two strange quarks. The mentioned models can reproduce $\Lambdac$ data better than $\Xiczeroplus$ data. This signals a possible difficulty with charm-strange baryons and suggests that the measurement of $\Omegac$ production represents a crucial step to constrain models and understand whether strange quarks, or strange diquarks, play a peculiar role in charm-baryon formation in pp collisions. 
In high-energy nucleus--nucleus collisions, the production yields of strange hadrons, in particular of multiple-strange baryons, normalised to pion ones are enhanced with respect to pp collisions and are well described by statistical models using a grand canonical ensemble with strangeness production regulated by chemical equilibrium~\cite{Rafelski:1982pu,WA97:1999uwz,STAR:2007cqw,Blume:2011sb,ALICE:2016fzo,ALICE:2013xmt,Andronic:2017pug}. Measurements of $\Omega^-$ and $\Xi^-$ production as a function of the event multiplicity suggest that the onset of such enhancement occurs progressively with increasing particle multiplicity, starting from low-multiplicity pp collisions~\cite{ALICE:2016fzo}. In this context, it is however interesting to note that although current data do not exclude that the $\Ds/\Dzero$ ratio in pp collisions could be larger than in \ee collisions, they do not support an increase similar, in relative terms, to that of $\Xiczeroplus/\Dzero$ ratio. Indeed, the analysis of charm fragmentation fractions reported in Ref.~\cite{ALICE:2021dhb} suggests that the sum of the ${\rm c}\rightarrow \Xiczero$ and ${\rm c} \rightarrow \Xicplus$  fragmentation fractions could be larger than the ${\rm c}\rightarrow \Ds$ one. Another interesting observation is given by the fact that the $\Xiczeroplus/\Sigmac$ ratio is described well by the default PYTHIA~8 Monash tune~\cite{ALICE:2021bli}, which significantly underestimates both $\Xiczeroplus/\Dzero$ and $\Sigmac/\Dzero$ ratios, suggesting that the production of the two baryons could be equally suppressed in \ee collisions because of similar mechanisms. The fraction of $\Lambdac$ coming from $\Sigmac$ decays is larger by a factor of about two in pp collisions than in \ee collisions~\cite{ALICE:2021rzj}: this supports the interpretation~\cite{Christiansen:2015yqa,Belle:2017caf}that in \ee collisions the $\Sigmac$ formation is suppressed by the need of forming in string breaking a (dd, ud, uu)-diquark with spin $S=1$, which is heavier than the $S=0$ (ud)-diquark needed to form a $\Lambdac$. A similar argument might be relevant in the comparison of $\Omegac$ and $\Xiczeroplus$ production, possibly influenced by the different mass values of $S=1$ (ss) and $S=0$ (sd, su) diquarks~\cite{Ebert:2011kk}. This further highlights the importance of measuring the $\Omegac$ production cross section to understand the role played by strange quarks and diquarks in charm-quark hadronisation. 
The measurement of the production cross section of the $\Omegac$ baryon is also needed to quantify its possible significant contribution to the total charm cross section at midrapidity per unit of rapidity, both in pp and in Pb--Pb collisions at the LHC~\cite{ALICE:2021dhb}.

This Letter reports on the first measurement of the $p_{\rm T}$-differential production cross section of the inclusive $\rm \Omega_c^0$ baryon multiplied by the branching ratio (BR) of the hadronic decay channel $\rm \Omega_c^0\rightarrow\Omega^-\pi^+$ at midrapidity ($|y|<0.5$) in pp collisions at $\sqrt{s}=13$~\TeV. Inclusive $\rm \Omega_c^0$ include prompt $\rm \Omega_c^0$, produced directly in the hadronisation of charm quarks or in the decay of directly produced excited charm states, as well as $\rm \Omega_c^0$ from decays of beauty or multiple-charm hadron decays. 
The ratios of the inclusive $\rm \Omega_c^0$ cross section to the prompt $\rm D^0$ meson~\cite{ALICE:2021rzj} and to the prompt charm--strange $\rm \Xi_c^0$ baryon~\cite{ALICE:2021bli} are also reported. The absolute branching ratio of the decay channel used has not been measured yet. The $\rm \Omegac$ baryon was reconstructed 
together with its charge conjugate in the interval $2<p_{\rm T}<12$ GeV$/c$.

A description of the ALICE detector and its performance can be found in Refs.~\cite{Aamodt:2008zz,Alice_performance}. The main detectors used for this measurement are the Inner Tracking System (ITS), the Time Projection Chamber (TPC), and the Time-Of-Flight detector (TOF). They are located in the central barrel, which covers the pseudorapidity interval ($|\eta|<0.9$), and are embedded in a solenoidal magnet that provides a $\rm B=0.5$~T field parallel to the beam direction. The ITS is used for tracking, vertex reconstruction, and trigger purposes. The TPC is the main tracking detector in the central barrel and is also used for particle identification (PID) via the measurement of the particle specific energy loss (${\rm d}E/{\rm d}x$). The TOF provides PID information via the measurement of the particle time-of-flight relative to the time of the collision~\cite{ALICE:2016ovj}. The analysed data sample consists of pp collisions at $\sqrt{s}=13$~TeV recorded with a minimum-bias (MB) trigger based on coincident signals in the two scintillator arrays (V0) located on both sides of the nominal interaction point along the beam direction. Offline selection criteria, based on the signals from the V0 and the Silicon Pixel Detector, which constitutes the two innermost ITS layers, were applied to remove background due to the interaction between one of the beams and the residual gas present in the beam vacuum tube as well as other machine-induced backgrounds~\cite{ALICE:2014sbx}. Events with multiple reconstructed primary vertices, which amount to 1\% of the total event sample, were rejected to reduce the contamination from the superposition of several collisions within the same colliding bunches (pile-up events). Only events with a primary vertex position within 10 cm from the nominal interaction point along the beam direction were used. 
After the aforementioned selections, the data sample corresponds to an integrated luminosity $\mathcal{L}_{\rm int}=$~32.08 $\pm$ 0.51~${\rm nb^{-1}}$~\cite{aliceLumi13TeVrun2}. 

\begin{sloppypar}
The $\rm \Omega_c^0$-baryon candidates were built from $\rm \Omega^-\pi^+$ pairs using a Kalman-Filter (KF) vertexing algorithm~\cite{KalmanFilter} by combining a positive charged track ($\rm \pi^+$ candidate) originating from the primary vertex and a $\rm \Omega^-$-baryon candidate. The $\rm \Omega^-$ was reconstructed from the decay chain $\rm \Omega^{-} \rightarrow \Lambda K^{-}$, ${{\rm BR}=(67.8\pm 0.7 )}$\%, followed by $\rm \Lambda \rightarrow p \pi^-$, BR $=$ ($63.9 \pm 0.5$)\%~\cite{ParticleDataGroup:2020ssz}. The $\rm \Omega^{-}$ and $\rm \Lambda$ baryons were reconstructed by exploiting their characteristic decay topologies as reported in Refs.~\cite{ALICE:2021bli,ALICE:2019avo}. The tracks of the charged particles involved in the decay chain were required to be in the pseudorapidity interval $|\eta|<0.8$, to have at least 70 out of 159 crossed TPC tracking points, and to have a fit quality $\chi^2/{\rm NDF}<2$ in the TPC.
Moreover, primary $\rm \pi^+$ candidates were required to have a minimum of four (out of six) hits in the ITS. Protons, pions, and kaons were selected by requiring compatibility within four standard deviations ($4\sigma$) between the measured signal and that expected for the respective particle hypothesis for both the TPC ${\rm d}E/{\rm d}x$ and the time-of-flight measurement. Tracks without signal in the TOF detector were identified using only the TPC information. 
In order to reduce the large combinatorial background, a machine-learning approach based on the adaptive Boosted Decision Tree (BDT) algorithm in the Toolkit for Multivariate Data Analysis (TMVA)~\cite{Hocker:2007ht} was used. The signal sample of $\rm \Omega_c^0$ baryons for the BDT training was obtained from a simulation based on the PYTHIA~8.243 event generator~\cite{Sjostrand:2014zea}. The mean proper lifetime of $\rm \Omega_c^0$ in the simulation was set to 80 $\rm \mu m$ according to the latest LHCb measurement~\cite{LHCb:2018nfa}. The propagation of the generated particles through the detector was performed using the GEANT 3 package~\cite{Brun:1082634}. The luminous region distribution and the conditions of all ALICE detectors in terms of active channels, gain, noise level, and alignment, and their evolution with time during the data taking, were taken into account in the simulations. The background candidates were taken from data by selecting
candidates with invariant mass in the intervals $2.39<M<2.62$~GeV$/c^2$ and $2.77<M<2.99$~GeV$/c^2$, which are outside of the expected mass peak of the $\rm \Omega_c^0$. Before the training, loose selections were applied on the distance, normalised to its uncertainty, between the $\rm \Lambda$ decay point and the primary vertex, 
and on the $\rm \Lambda$, $\rm \Omega^-$, and $\rm \Omega_c^0$ $\chi^2_{\rm geo}/\rm NDF$, which is a variable calculated by the KF Particle algorithm~\cite{KalmanFilter} related to the intersection probability of the daughter-particle trajectories taking their uncertainties into account. The BDT model was trained independently for each $p_{\rm T}$ interval with variables related to the $\rm \Omega^-$ decay topology, such as the distance of closest approach (DCA) of the decay particles, the DCA between the primary vertex and the reconstructed $\rm \Omega^-$ candidate, the pointing angle of the reconstructed $\rm \Omega^-$ decay vertex to the reconstructed $\rm \Omega_c^0$ decay vertex, the $\chi^2_{\rm geo}/\rm NDF$, and the $\chi^2_{\rm topo}/\rm NDF$. The $\chi^2_{\rm topo}/\rm NDF$ is calculated by the KF Particle~\cite{KalmanFilter} algorithm and characterises whether the $\rm \Omega^-$ candidate points back to the reconstructed $\rm \Omega_c^0$ decay vertex. 
The output of the BDT training allows the classification of each candidate with a number related to its probability to be a $\rm \Omega_c^0$ baryon signal or combinatorial background. 
\end{sloppypar}

\begin{figure}[ht!]
    \begin{center}
    \includegraphics[width = 0.49\textwidth]{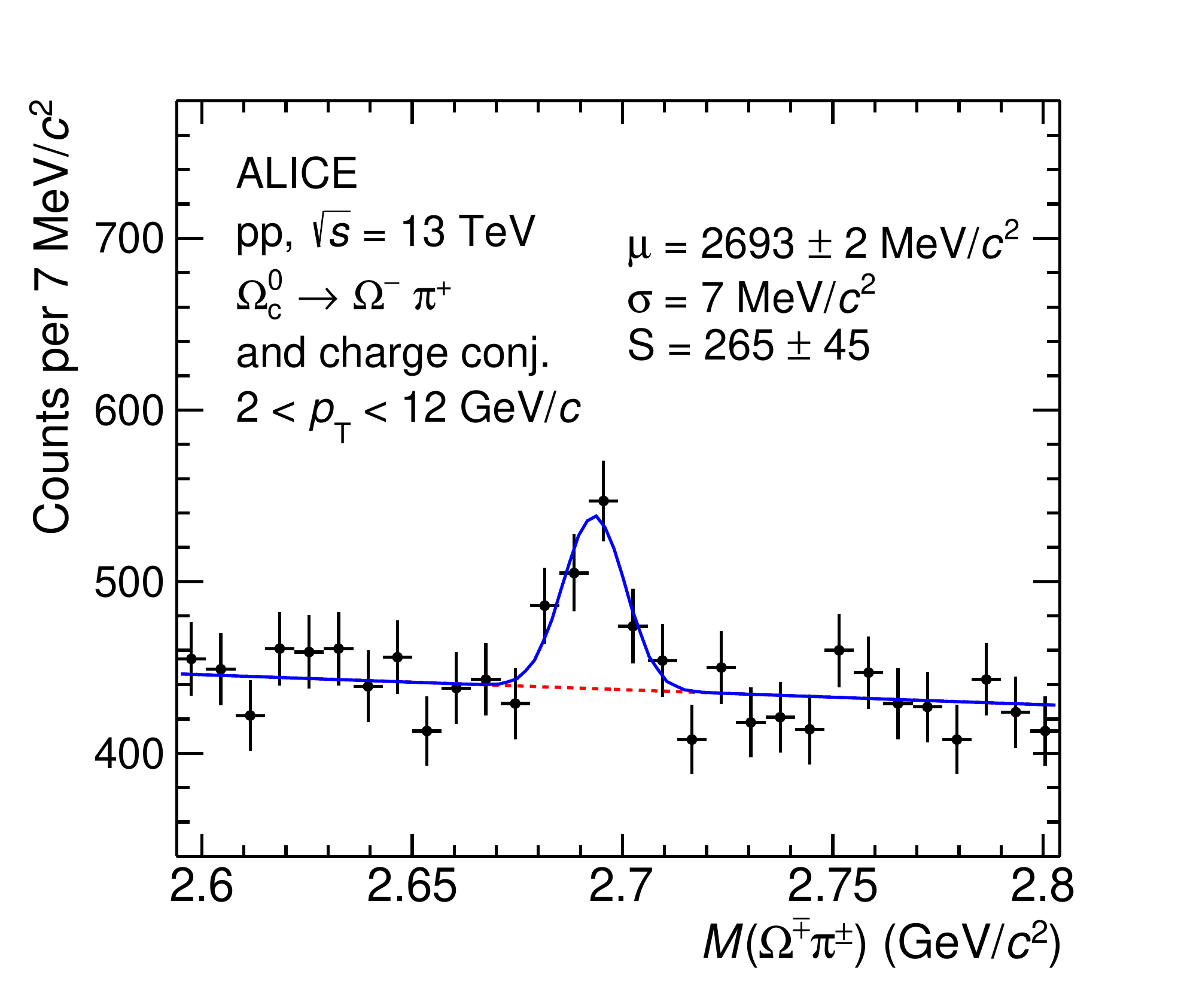}
       \includegraphics[width = 0.49\textwidth]{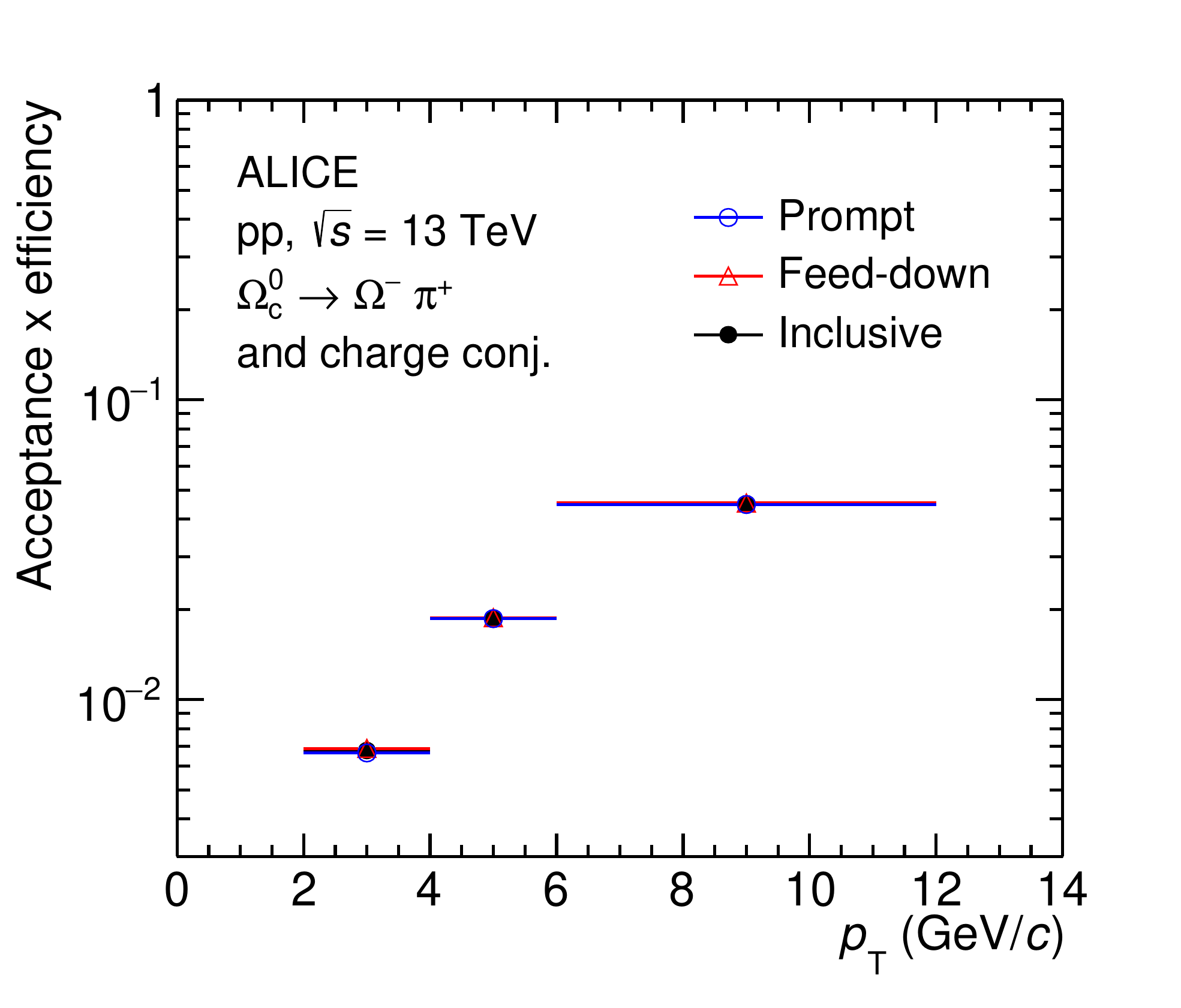}
    \end{center}
    \caption{(Left panel): invariant-mass distribution of $\rm \Omega_c^0\rightarrow\Omega^-\pi^+$ candidates and their charge conjugates  integrated over the whole \pt interval 2--12 GeV$/c$. The blue line shows the total fit function and the red line represents the combinatorial background fit. (Right panel): acceptance-times-efficiency for prompt, feed-down, and inclusive $\rm \Omega_c^0$ baryons decaying into $\rm \Omega^-\pi^+$ as a function of $p_{\rm T}$ in pp collisions at $\sqrt{s}=13$~TeV.}
    \label{fig:InvMass_PtIntegral}
\end{figure}

The $\rm \Omega_c^0$ raw yields were obtained from the fit to the invariant-mass distribution of the candidates as shown in the left panel of Fig.~\ref{fig:InvMass_PtIntegral}. The signal peak was modelled with a Gaussian function and the background was described by a linear function. 

The $p_{\rm T}$ and $y$-differential production cross section in the rapidity interval $|y|<0.5$ of inclusive $\rm \Omega_c^0$ baryons multiplied by the branching ratio into the considered hadronic decay channel was calculated from the raw yields as follows

\begin{equation}
	{\rm BR}\times\frac{\text{d}^2\sigma^{\Omega^0_{\rm c}}}{\text{d}p_\mathrm{T}\text{d}y}=\frac{1}{2\Delta y\Delta p_\mathrm{T}}\times\frac{N^{\Omega^{0}_{\rm c}+\overline{\Omega}{}^{0}_{\rm c}}_\text{raw}}{({\rm Acc} \times{\it \varepsilon})_{\rm inclusive}}\times\frac{1}{\mathcal{L}_{\rm int}},
	\label{eq:CrossSection}
\end{equation}

where $N^{\Omega^{0}_{\rm c}+\overline{\Omega}{}^{0}_{\rm c}}_\text{raw}$ is the raw yield in a given $p_{\rm T}$ interval with width $\Delta p_{\rm T}$ and in the rapidity interval $\Delta y=1.6$ assuming that the cross section does not vary significantly from $|y|<0.5$ to $|y|<0.8$. To confirm that this assumption has a negligible impact on the result, it was verified that by assuming the rapidity dependence expected for charm mesons in FONLL~\cite{Cacciari_1998,Cacciari_2012} and for charm baryons in PYTHIA~8~\cite{Sjostrand:2014zea} the cross section changes by less than $1\%$ in the measured $\pt$ interval. Since the feed-down contribution is not subtracted, the raw yield is divided by the inclusive acceptance-times-efficiency factor, $(\rm Acc \times {\it \varepsilon})_{inclusive}$ and by the integrated luminosity $\mathcal{L}_{\rm int}$ of the data sample to obtain the production cross section. The factor $1/2$ is needed to compute the average cross section of $\rm \Omega_c^0$ and $\rm \overline{\Omega}_c^0$. 
The factor $({\rm Acc} \times{\it \varepsilon})_{\rm inclusive}$ is the product of the geometrical acceptance (Acc) and the reconstruction and selection efficiency ($\varepsilon$) for the $\rm \Omega_c^0\rightarrow\Omega^-\pi^+$ decay.
The $({\rm Acc} \times{\it \varepsilon})_{\rm inclusive}$ correction was obtained from a simulation with the same configuration as
the one used for the BDT training described above.
The $\rm \Omega_c^0$-baryon \pt distribution from the simulations was reweighted in order to use realistic momentum distributions in the determination of the acceptance and the efficiency, which depends on \pt. The weights were defined with an iterative procedure to match the \pt dependence measured for $\rm \Omega_c^0$ baryon in the intervals used in the analysis. In the simulation, the $\rm \Omega_c^0$ is unpolarised: it was assumed that the modification of the acceptance that would arise from a non-zero polarisation can be considered negligible with respect to the statistical uncertainty and the other systematic uncertainties of the measurement. 
The right panel of Fig.~\ref{fig:InvMass_PtIntegral} shows the final $(\rm Acc \times {\it \varepsilon})$ correction factors of prompt, beauty feed-down, and inclusive $\rm \Omega_c^0$ as a function of $p_{\rm T}$. They are consistent with each other within uncertainties because 
the selection variables used are not sensitive to the displacement by a few hundred micrometers
of the prompt and beauty feed-down $\rm \Omega_c^0$ decay vertices from the collision point. The efficiency values increase with $\pt$ from about 0.7\% to about 5\%.

Systematic uncertainties were estimated considering several sources.
The uncertainty on the track
reconstruction efficiency was evaluated by varying the track selection criteria and by comparing the probability to prolong the tracks from the TPC to the ITS hits in data and simulations. A 6\% uncertainty was assigned. The systematic uncertainty on the selection efficiency derives from possible differences between the detector resolutions and alignment and their description in the simulation. This uncertainty was assessed from the comparison of the corrected yields obtained by varying the selections. In particular, the selections on the BDT outputs 
were varied separately in the different $p_{\rm T}$ intervals, with a corresponding variation of the  efficiencies ranging from 30\% to 50\% depending on $p_{\rm T}$. The assigned systematic uncertainty is 10\%, which represents the largest contribution to the systematic uncertainty of the measurement. The systematic uncertainty due to the shape of the $\rm \Omega_c^0$ $p_{\rm T}$ spectrum used in the simulation for the calculation of the $(\rm Acc \times {\it \varepsilon})_{inclusive}$ factor was estimated by modifying the weights mentioned above within their uncertainties. An uncertainty of about 4\% was estimated in the \pt interval $2<\pt<4$~GeV$/c$ and a 2\% uncertainty in $4<\pt<12$~GeV$/c$. The systematic uncertainty on the raw-yield extraction was evaluated in each \pt interval by repeating the fit to the invariant-mass distributions varying the function used to describe the background and the fit range. 
In order to test the sensitivity to the line-shape of the signal, a bin-counting method was used, in which the signal yield was obtained by integrating the invariant-mass distribution after subtracting the combinatorial background. A 6\% uncertainty was assigned independent of $p_{\rm T}$. 
The sources of systematic uncertainty are assumed to be uncorrelated among each other
and the total systematic uncertainty in each \pt interval is calculated by a quadratic
sum of the individual contributions, resulting in a 14\% systematic uncertainty in $2<p_{\rm T}<4$ GeV$/c$ and 13\% in $4<p_{\rm T}<12$ GeV$/c$. The production cross section has an additional global normalisation uncertainty of 1.6\% due to the integrated luminosity determination~\cite{aliceLumi13TeVrun2}. 

\begin{figure}[ht!]
    \begin{center}
    \includegraphics[width = 0.6\textwidth]{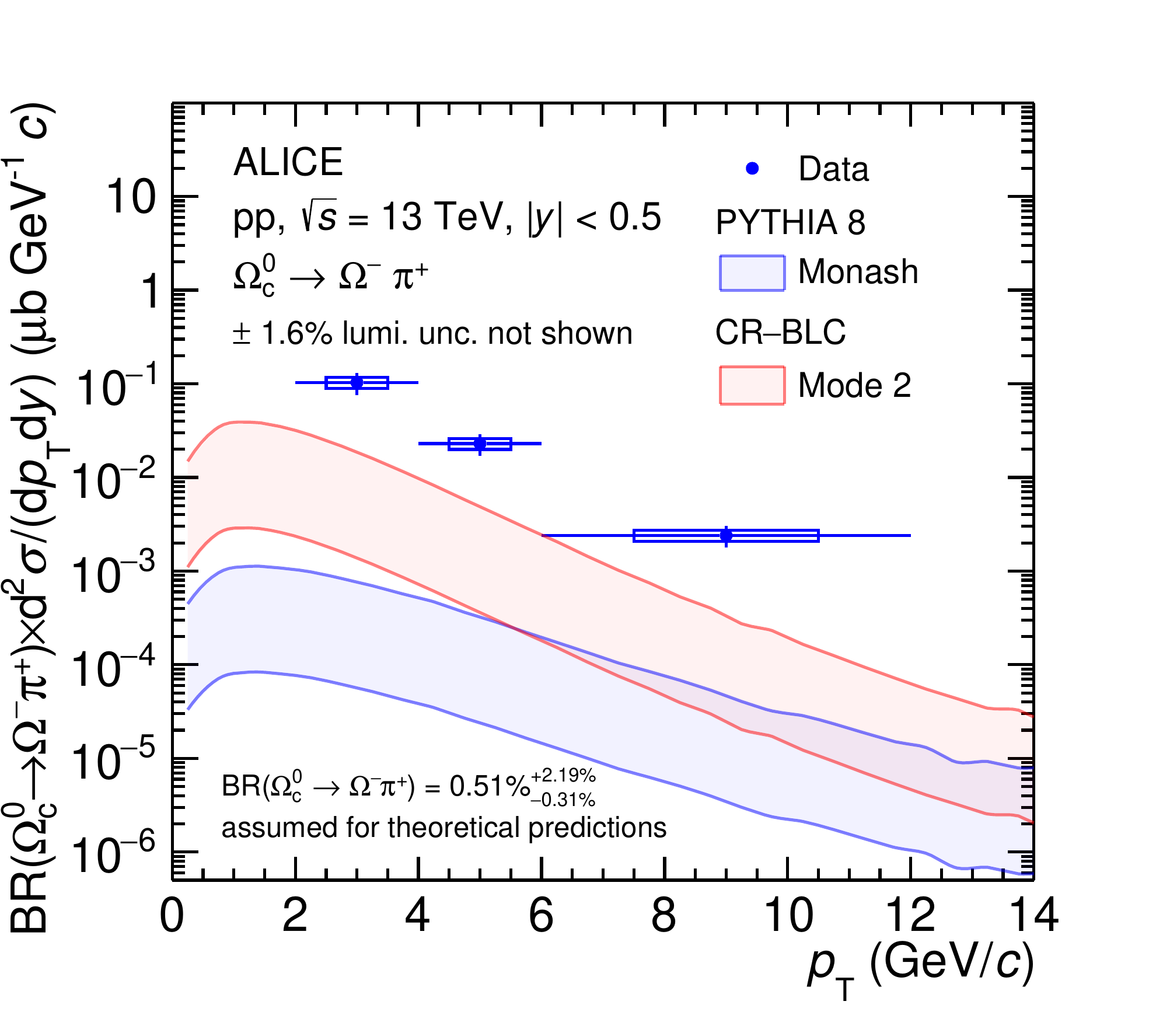}
    \end{center}
    \caption{The $p_{\rm T}$-differential production cross section of inclusive $\rm \Omega_c^0$ baryons multiplied by the branching ratio into $\rm \Omega^-\pi^+$ for $|y|<0.5$ in pp collisions at $\sqrt{s}=13$~TeV. The error bars and empty boxes represent the statistical and systematic uncertainties, respectively. The measurement is compared with PYTHIA~8.243 with Monash tune~\cite{Skands:2014pea} and with CR beyond the leading-colour approximation~\cite{Christiansen:2015yqa}, which are multiplied by a theoretical $\rm BR(\Omega_c^0\rightarrow\Omega^-\pi^+)=(0.51^{+2.19}_{-0.31})\%$~\cite{Hsiao:2020gtc, Gutsche:2018utw, Cheng:1996cs, Hu:2020nkg, Solovieva:2008fw, Wang:2022zja}.}
    \label{fig:BRxCR}
\end{figure}

The $p_{\rm T}$-differential production cross section of inclusive $\rm \Omega_c^0$ baryons multiplied by the branching ratio of the $\rm \Omega^-\pi^+$ channel measured in the rapidity interval $|y|<0.5$ and the $p_{\rm T}$ interval $2<p_{\rm T}<12$~GeV$/c$ are shown in Fig.~\ref{fig:BRxCR}. The feed-down contribution from $\rm \Omega_b^-$, e.g. $\rm \Omega_b^- \rightarrow \rm \Omega_c^0 + \pi^-$~\cite{ParticleDataGroup:2020ssz}, is not subtracted because of the lack of knowledge of the  branching ratios of b-hadron decays to $\rm \Omegac$. Given that the efficiencies of prompt and feed-down $\Omegac$ are consistent within uncertainties, the inclusive measurement presented here preserve the original relative abundances of its prompt and feed-down components. The data are compared with the inclusive $\Omegac$ $\pt$-differential cross sections expected from the PYTHIA~8.243 Monash and CR-BLC tunes (Mode 2)~\cite{Sjostrand:2014zea, Skands:2014pea, Christiansen:2015yqa} multiplied by the branching ratio, $\rm BR(\Omega_c^0\rightarrow\Omega^-\pi^+)=(0.51^{+2.19}_{-0.31})\%$, obtained by considering the estimate reported in Ref.~\cite{Hsiao:2020gtc} for the central value, and the envelope of the values (including their uncertainties) reported in Refs.~\cite{Hsiao:2020gtc, Gutsche:2018utw, Cheng:1996cs, Hu:2020nkg, Solovieva:2008fw, Wang:2022zja} to determine the uncertainty. 
In the $p_{\rm T}$ interval of the measurement, the cross section from the CR-BLC tune is larger than the one from the Monash tune by factor varying between 9 and 25 depending on $p_{\rm T}$. The Monash tune and CR-BLC tune underestimate the data by more than 3.3$\sigma$ and 2.7$\sigma$, respectively, when $\rm BR(\Omega_c^0\rightarrow\Omega^-\pi^+)=0.51\% ^{+2.19\%} _{-0.31\%}$ is considered.

\begin{figure}[ht!]
\centering
\includegraphics[width=.49\textwidth]{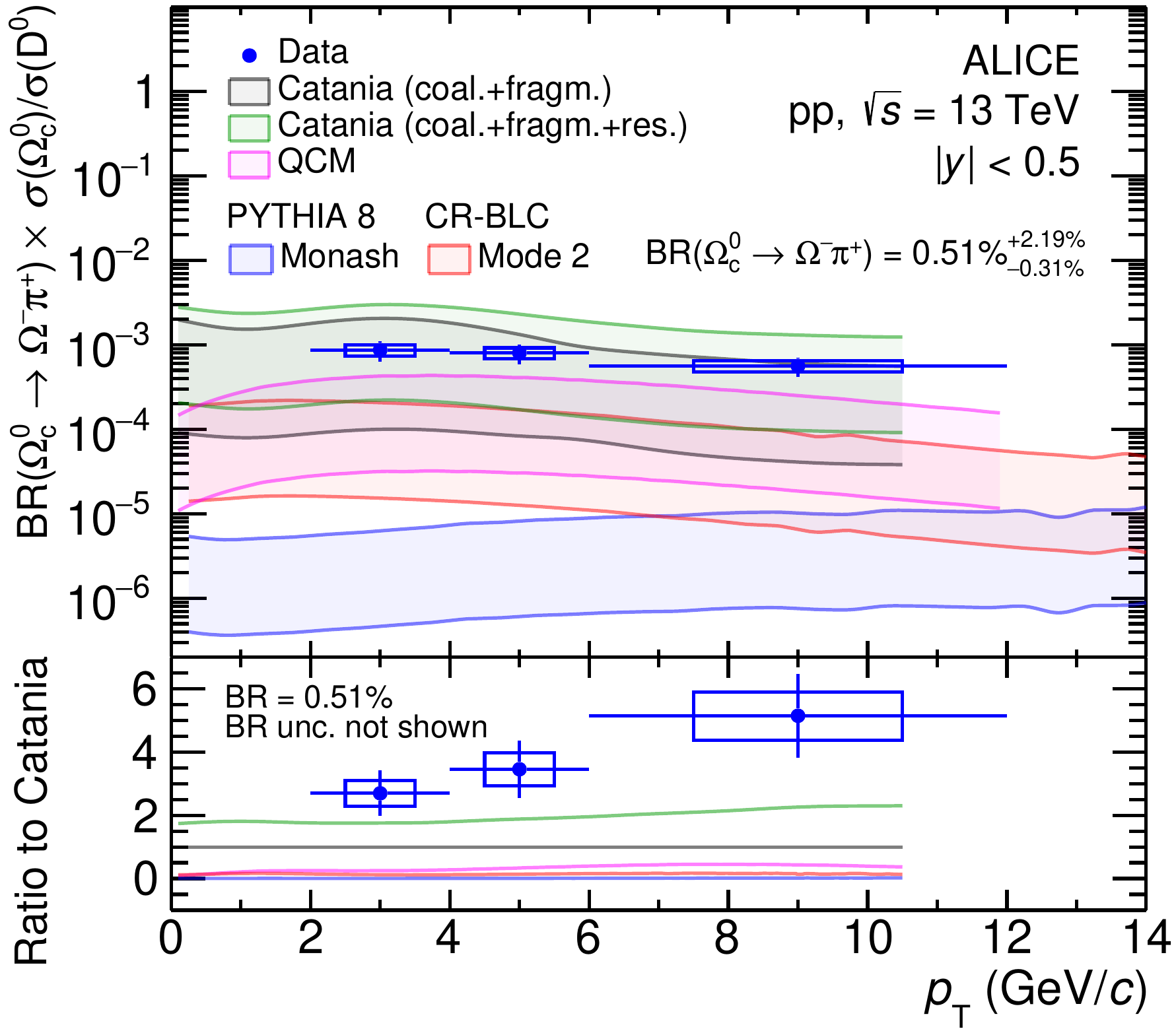}
\includegraphics[width=.49\textwidth]{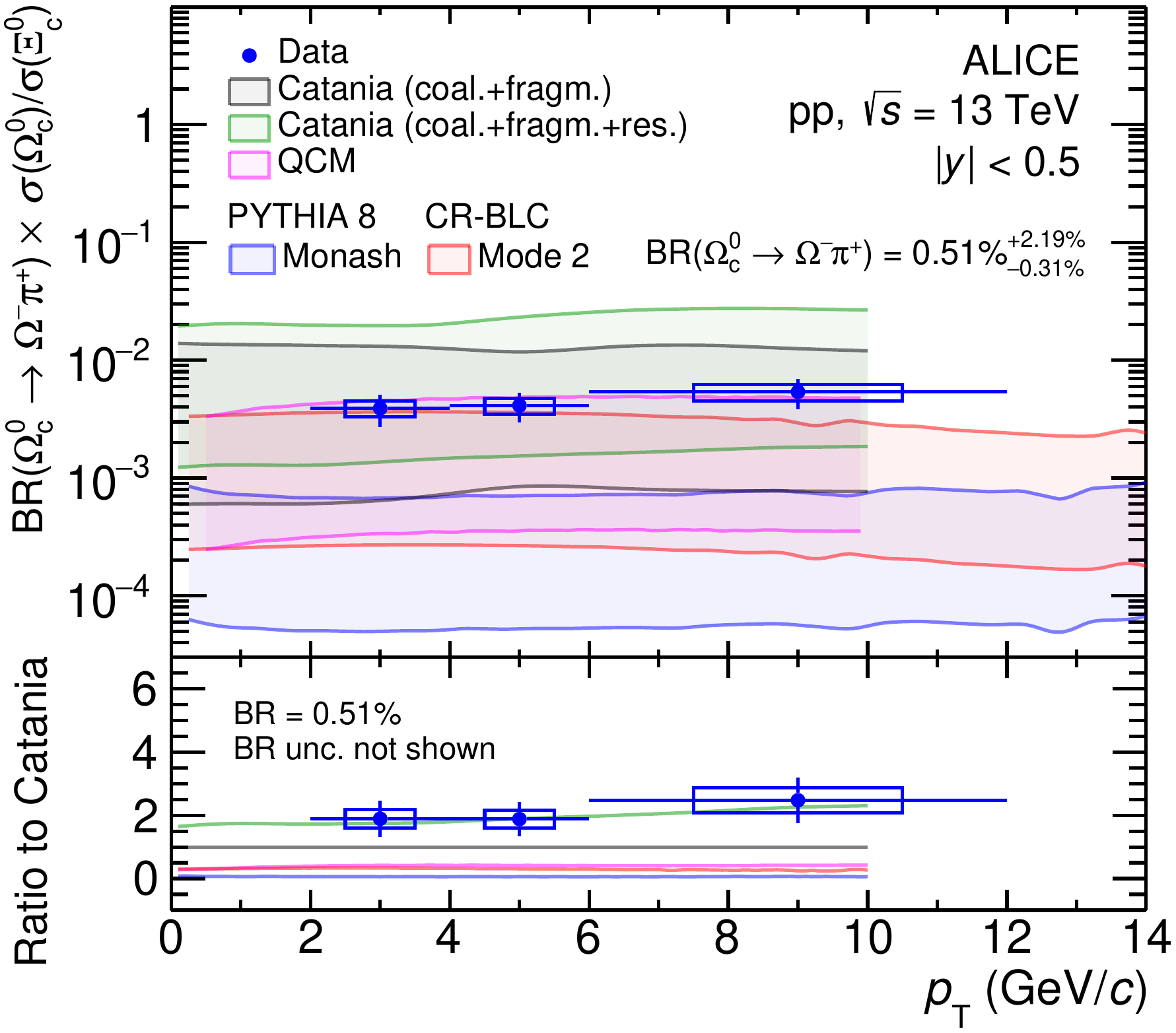}
\caption{Left, top panel: ratio of the $p_{\rm T}$-differential cross section of $\rm \Omega_c^0$ baryons (multiplied by the branching ratio into $\rm \Omega^-\pi^+$) to the $\rm D^0$-meson one~\cite{ALICE:2021rzj} in $|y|<0.5$ in pp collisions at $\sqrt{s}=13$~TeV. Right, top panel: ratio of the $p_{\rm T}$-differential cross section of $\rm \Omegac$ baryons (multiplied by the branching ratio into $\rm \Omega^-\pi^+$) to the $\rm \Xiczero$-baryon one~\cite{ALICE:2021bli} in $|y|<0.5$ in pp collisions at $\sqrt{s}=13$~TeV. Bottom panels: ratio of the data and models to the Catania (coalescence plus fragmentation) model~\cite{Minissale:2020bif}. The error bars and empty boxes represent the statistical and systematic uncertainties, respectively. The measurements are compared with model calculations (see text for details), which are multiplied by a theoretical $\rm BR(\Omega_c^0\rightarrow\Omega^-\pi^+)=(0.51^{+2.19}_{-0.31})\%$~\cite{Hsiao:2020gtc, Gutsche:2018utw, Cheng:1996cs, Hu:2020nkg, Solovieva:2008fw, Wang:2022zja}.}
\label{fig:BaryonToMesonAndToBaryon}
\end{figure}

The ratios of the $p_{\rm T}$-differential production cross section of inclusive $\rm \Omega_c^0$ baryons (multiplied by the branching ratio of the $\Omegac \rightarrow \Omega^-\pi^+$ decay channel) to the prompt $\rm D^0$-meson cross section~\cite{ALICE:2021rzj} and to the prompt $\rm \Xiczero$-baryon one~\cite{ALICE:2021bli} 
are reported in the left and right panel of Fig.~\ref{fig:BaryonToMesonAndToBaryon}, respectively.  The systematic uncertainties on the tracking efficiency and on the luminosity were propagated as fully correlated in the ratios. The uncertainties do not allow to draw a conclusion about the possible \pt dependence of the ratios. The data are compared with model expectations that were obtained by scaling the $\Omegac/\Dzero$ and $\rm \Omegac/\Xi^0_c$ ratios predicted by the models by the BR of the $\rm \Omega_c^0\rightarrow\Omega^-\pi^+$ decay channel mentioned above. The uncertainty band of the models represents the BR uncertainty. For the Catania model only the specific uncertainty of the model itself are also included in the uncertainty band~\cite{Minissale:2020bif}. In the bottom panels, the ratios of the various models and the data to the Catania prediction are shown. The expectations of the models differ significantly, even by orders of magnitude, demonstrating the sensitivity of the measured ratios to the implementation of the charm hadronisation process in the models. As visible in the left panels of Fig.~\ref{fig:BaryonToMesonAndToBaryon}, the Monash~\cite{Skands:2014pea} and CR-BLC~\cite{Christiansen:2015yqa} tunes of PYTHIA~8, as well as the QCM~\cite{Song:2018tpv} model underestimate the data significantly. The Monash tune expects a $\rm BR(\Omegac \rightarrow \Omega^-\pi^+)\times \Omegac/\Dzero$ ratio increasing with $\pt$ from about $4\times10^{-7}$ to about $1\times10^{-5}$. The CR-BLC model enhances the ratio by a factor of 12 to 34 with respect to the Monash tune. The prediction of the QCM is larger than that of the CR-BLC model, but it is lower than the data by more than 1.8$\sigma$. The Catania model~\cite{Minissale:2020bif} is consistent with the data. In particular, in the version in which additional charm resonance states on top of those listed in the PDG~\cite{ParticleDataGroup:2020ssz} are considered, the $\Omegac/\Dzero$ ratio is enhanced by a factor of 2, thus enlarging the range of possible BR($\Omegac \rightarrow \Omega^-\pi^+$) values that would allow the model prediction to be compatible within $1\sigma$ with the data considering only the data uncertainty. 
The $\Omegac/\Dzero$ ratio decreases with $\pt$ in the measured $\pt$ range in the CR-BLC, QCM, and Catania models, oppositely to what is expected by Monash. 
In the $\Omegac/\Xiczero$ baryon-to-baryon ratio, shown in the right panel of Fig.~\ref{fig:BaryonToMesonAndToBaryon}, a similar hierarchy among the model predictions is present, though PYTHIA~8 with CR-BLC gives an enhancement by a factor of 4 to 5 with respect to the the Monash expectation, thus smaller than that of the $\Omegac/\Dzero$ ratio. Also for this ratio, the CR-BLC and QCM predictions are close to each other and higher than the Monash tune. 
The Catania model shows a good agreement with the data, whether the augmented set of charm resonance states is considered or not.

Using the ALICE $\Xiczero$~\cite{ALICE:2021bli} and $\Lambdac$~\cite{ALICE:2021rzj} data, the ratios $\rm BR(\Omega_c^0\rightarrow\Omega^-\pi^+)\times{\it \sigma}(\Omega_c^0)/{\it \sigma}(\Lambda_c^+)$ and $\rm BR(\Omega_c^0\rightarrow\Omega^-\pi^+)\times{\it \sigma}(\Omega_c^0)/{\it \sigma}(\Xi_c^0)$ of the cross sections integrated in the $\Omegac$ measured \pt interval were obtained. They are reported in 
Table~\ref{tab:ALICEvsBelle}. They are compared with the values measured in $\rm e^+e^-$ collisions at $\sqrt{s}=10.52$~GeV by Belle, obtained from the cross sections reported in Table~1 of Ref.~\cite{Belle:2017caf}. Though the limited $\pt$ and rapidity ranges of the ALICE measurement do not allow for a direct comparison of the pp and $\ee$ data, the ratios observed by ALICE are larger by a factor of $8.7\pm2.2{\rm (stat.)}\pm0.9{\rm (syst.)}$ and $4.7\pm1.3{\rm (stat.)}\pm0.5{\rm (syst.)}$ for the $\rm BR(\Omega_c^0\rightarrow\Omega^-\pi^+)\times{\it \sigma}(\Omega_c^0)/{\it \sigma}(\Lambda_c^+)$ and $\rm BR(\Omega_c^0\rightarrow\Omega^-\pi^+)\times{\it \sigma}(\Omega_c^0)/{\it \sigma}(\Xi_c^0)$, respectively. The large BR uncertainties of the ${\rm \Xi_c^0}$ are not propagated in the computation of this factor. This difference, along with the comparison of data and  models in Fig.~\ref{fig:BaryonToMesonAndToBaryon}, represents further evidence that the hadronisation process differs in pp and $\ee$ collisions and is sensitive to the density of quarks, colour charges, and on the system size.

\begin{table}[ht!]
\caption{Ratio of the $p_{\rm T}$-integrated cross section of $\rm \Omega_c^0$ baryon (multiplied by the branching ratio into $\rm \Omega^-\pi^+$) in the interval $2<\pt<12$~\GeVc with respect to the $\rm \Lambda_c^+$- and $\rm \Xi_c^0$-baryon cross sections measured by the ALICE~\cite{ALICE:2021rzj,ALICE:2021bli} and Belle~\cite{Belle:2017caf} experiments in pp collisions at $\sqrt{s}=13$~\TeV and $\ee$ collisions at $\sqrt{s}=10.52$~\GeV, respectively. The first and second uncertainties represent the statistical and systematic ones. The data include the correction for the branching ratio BR($\Omega^-\rightarrow\Lambda{\rm K^-},\Lambda\rightarrow {\rm p}\pi^{-}$)=$(43.3\pm 0.6)\%$~\cite{ParticleDataGroup:2020ssz}.}

	\begin{center}
	\renewcommand\arraystretch{1.2}
	\begin{tabular}{ccc}
	\hline
	\hline
	  Ratio & ALICE (pp 13~TeV) & Belle ($\rm e^+e^-$ 10.52~GeV)~\cite{Belle:2017caf} \\
	  & $2<p_{\rm T}<12$~GeV$/c$ & visible \\
	\hline
	  $\rm BR(\Omega_c^0\rightarrow\Omega^-\pi^+)\times{\it \sigma}(\Omega_c^0)/{\it \sigma}(\Lambda_c^+)$ & $(1.96\pm0.42\pm0.13)\times10^{-3}$ & $(2.24\pm0.29\pm0.16)\times10^{-4}$ \\
	  $\rm BR(\Omega_c^0\rightarrow\Omega^-\pi^+)\times{\it \sigma}(\Omega_c^0)/{\it \sigma}(\Xi_c^0)$ & $(3.99\pm0.96\pm0.96)\times10^{-3}$ & $(8.58\pm1.15\pm1.98)\times10^{-4}$ \\
	\hline
    \hline
	\end{tabular}
	\end{center}
	\label{tab:ALICEvsBelle}
\end{table}

In summary, the inclusive \pt-differential production cross section of the charm-strange
baryon $\rm \Omega_c^0$ multiplied by the branching ratio of the $\rm \Omega_c^0 \rightarrow \Omega^-\pi^+$ decay channel was measured at midrapidity ($|y|<0.5$) in pp collisions at $\sqrt{s}=13$~TeV. The ratio of this measurement to the production cross section of the $\Dzero$ meson provides further evidence that charm quarks hadronise to $\rm \Omega_c^0$ baryons more frequently in pp collisions than in $\ee$ collisions, confirming the general trend observed from previous measurements of $\Lambdac$, $\Xiczeroplus$, and $\Sigmac$ production. 
The large uncertainty of the $\Omegac\rightarrow\Omega^-\pi^+$ branching ratio limits the effectiveness of the comparison with theoretical models. However, the predictions of the available models differ by large factors indicating that future measurements of the BR, which could be performed also by the LHCb or Belle 2 collaborations, will allow to exploit these data to set stringent constraints to theoretical models and obtain deep insight into the charm hadronisation and the role of strange quarks and diquarks. Moreover, despite the large uncertainties, only the Catania model, which assumes that charm-quark hadronisation proceeds via both fragmentation and coalescence, can describe the $\rm BR(\Omega_c^0\rightarrow\Omega^-\pi^+)\times{\it \sigma}(\Omega_c^0)/{\it \sigma}(\Dzero)$ ratio within uncertainties. More precise
measurements with the data sample collected in Run 3 of the LHC will allow us to further investigate the
\pt shape of the $\Omegac/\Dzero$ and $\Omegac/\Xiczero$ ratios.


\newenvironment{acknowledgement}{\relax}{\relax}
\begin{acknowledgement}
\section*{Acknowledgements}

The ALICE Collaboration would like to thank all its engineers and technicians for their invaluable contributions to the construction of the experiment and the CERN accelerator teams for the outstanding performance of the LHC complex.
The ALICE Collaboration gratefully acknowledges the resources and support provided by all Grid centres and the Worldwide LHC Computing Grid (WLCG) collaboration.
The ALICE Collaboration acknowledges the following funding agencies for their support in building and running the ALICE detector:
A. I. Alikhanyan National Science Laboratory (Yerevan Physics Institute) Foundation (ANSL), State Committee of Science and World Federation of Scientists (WFS), Armenia;
Austrian Academy of Sciences, Austrian Science Fund (FWF): [M 2467-N36] and Nationalstiftung f\"{u}r Forschung, Technologie und Entwicklung, Austria;
Ministry of Communications and High Technologies, National Nuclear Research Center, Azerbaijan;
Conselho Nacional de Desenvolvimento Cient\'{\i}fico e Tecnol\'{o}gico (CNPq), Financiadora de Estudos e Projetos (Finep), Funda\c{c}\~{a}o de Amparo \`{a} Pesquisa do Estado de S\~{a}o Paulo (FAPESP) and Universidade Federal do Rio Grande do Sul (UFRGS), Brazil;
Bulgarian Ministry of Education and Science, within the National Roadmap for Research Infrastructures 2020-2027 (object CERN), Bulgaria;
Ministry of Education of China (MOEC) , Ministry of Science \& Technology of China (MSTC) and National Natural Science Foundation of China (NSFC), China;
Ministry of Science and Education and Croatian Science Foundation, Croatia;
Centro de Aplicaciones Tecnol\'{o}gicas y Desarrollo Nuclear (CEADEN), Cubaenerg\'{\i}a, Cuba;
Ministry of Education, Youth and Sports of the Czech Republic, Czech Republic;
The Danish Council for Independent Research | Natural Sciences, the VILLUM FONDEN and Danish National Research Foundation (DNRF), Denmark;
Helsinki Institute of Physics (HIP), Finland;
Commissariat \`{a} l'Energie Atomique (CEA) and Institut National de Physique Nucl\'{e}aire et de Physique des Particules (IN2P3) and Centre National de la Recherche Scientifique (CNRS), France;
Bundesministerium f\"{u}r Bildung und Forschung (BMBF) and GSI Helmholtzzentrum f\"{u}r Schwerionenforschung GmbH, Germany;
General Secretariat for Research and Technology, Ministry of Education, Research and Religions, Greece;
National Research, Development and Innovation Office, Hungary;
Department of Atomic Energy Government of India (DAE), Department of Science and Technology, Government of India (DST), University Grants Commission, Government of India (UGC) and Council of Scientific and Industrial Research (CSIR), India;
National Research and Innovation Agency - BRIN, Indonesia;
Istituto Nazionale di Fisica Nucleare (INFN), Italy;
Japanese Ministry of Education, Culture, Sports, Science and Technology (MEXT) and Japan Society for the Promotion of Science (JSPS) KAKENHI, Japan;
Consejo Nacional de Ciencia (CONACYT) y Tecnolog\'{i}a, through Fondo de Cooperaci\'{o}n Internacional en Ciencia y Tecnolog\'{i}a (FONCICYT) and Direcci\'{o}n General de Asuntos del Personal Academico (DGAPA), Mexico;
Nederlandse Organisatie voor Wetenschappelijk Onderzoek (NWO), Netherlands;
The Research Council of Norway, Norway;
Commission on Science and Technology for Sustainable Development in the South (COMSATS), Pakistan;
Pontificia Universidad Cat\'{o}lica del Per\'{u}, Peru;
Ministry of Education and Science, National Science Centre and WUT ID-UB, Poland;
Korea Institute of Science and Technology Information and National Research Foundation of Korea (NRF), Republic of Korea;
Ministry of Education and Scientific Research, Institute of Atomic Physics, Ministry of Research and Innovation and Institute of Atomic Physics and University Politehnica of Bucharest, Romania;
Ministry of Education, Science, Research and Sport of the Slovak Republic, Slovakia;
National Research Foundation of South Africa, South Africa;
Swedish Research Council (VR) and Knut \& Alice Wallenberg Foundation (KAW), Sweden;
European Organization for Nuclear Research, Switzerland;
Suranaree University of Technology (SUT), National Science and Technology Development Agency (NSTDA) and National Science, Research and Innovation Fund (NSRF via PMU-B B05F650021), Thailand;
Turkish Energy, Nuclear and Mineral Research Agency (TENMAK), Turkey;
National Academy of  Sciences of Ukraine, Ukraine;
Science and Technology Facilities Council (STFC), United Kingdom;
National Science Foundation of the United States of America (NSF) and United States Department of Energy, Office of Nuclear Physics (DOE NP), United States of America.
In addition, individual groups or members have received support from:
Marie Sk\l{}odowska Curie, European Research Council, Strong 2020 - Horizon 2020 (grant nos. 950692, 824093, 896850), European Union;
Academy of Finland (Center of Excellence in Quark Matter) (grant nos. 346327, 346328), Finland;
Programa de Apoyos para la Superaci\'{o}n del Personal Acad\'{e}mico, UNAM, Mexico.

\end{acknowledgement}

\bibliographystyle{utphys}   
\bibliography{bibliography}

\newpage
\appendix

%
%

\section{The ALICE Collaboration}
\label{app:collab}
\begin{flushleft} 
\small

S.~Acharya\,\orcidlink{0000-0002-9213-5329}\,$^{\rm 124,132}$, 
D.~Adamov\'{a}\,\orcidlink{0000-0002-0504-7428}\,$^{\rm 85}$, 
A.~Adler$^{\rm 69}$, 
G.~Aglieri Rinella\,\orcidlink{0000-0002-9611-3696}\,$^{\rm 32}$, 
M.~Agnello\,\orcidlink{0000-0002-0760-5075}\,$^{\rm 29}$, 
N.~Agrawal\,\orcidlink{0000-0003-0348-9836}\,$^{\rm 50}$, 
Z.~Ahammed\,\orcidlink{0000-0001-5241-7412}\,$^{\rm 132}$, 
S.~Ahmad\,\orcidlink{0000-0003-0497-5705}\,$^{\rm 15}$, 
S.U.~Ahn\,\orcidlink{0000-0001-8847-489X}\,$^{\rm 70}$, 
I.~Ahuja\,\orcidlink{0000-0002-4417-1392}\,$^{\rm 37}$
% ...

\section*{Affiliation Notes}

$^{\rm I}$ Deceased\\
$^{\rm II}$ Also at: Max-Planck-Institut f\"{u}r Physik, Munich, Germany\\
$^{\rm III}$ Also at: Italian National Agency for New Technologies, Energy and Sustainable Economic Development (ENEA), Bologna, Italy\\
$^{\rm IV}$ Also at: Dipartimento DET del Politecnico di Torino, Turin, Italy\\
$^{\rm V}$ Also at: Department of Applied Physics, Aligarh Muslim University, Aligarh, India\\
$^{\rm VI}$ Also at: Institute of Theoretical Physics, University of Wroclaw, Poland\\
$^{\rm VII}$ Also at: An institution covered by a cooperation agreement with CERN\\

\section*{Collaboration Institutes}

$^{1}$ A.I. Alikhanyan National Science Laboratory (Yerevan Physics Institute) Foundation, Yerevan, Armenia\\
$^{2}$ AGH University of Krakow, Cracow, Poland\\
$^{3}$ Bogolyubov Institute for Theoretical Physics, National Academy of Sciences of Ukraine, Kiev, Ukraine\\
$^{4}$ Bose Institute, Department of Physics  and Centre for Astroparticle Physics and Space Science (CAPSS), Kolkata, India\\
$^{5}$ California Polytechnic State University, San Luis Obispo, California, United States\\
$^{6}$ Central China Normal University, Wuhan, China\\
$^{7}$ Centro de Aplicaciones Tecnol\'{o}gicas y Desarrollo Nuclear (CEADEN), Havana, Cuba\\
$^{8}$ Centro de Investigaci\'{o}n y de Estudios Avanzados (CINVESTAV), Mexico City and M\'{e}rida, Mexico\\
$^{9}$ Chicago State University, Chicago, Illinois, United States\\
$^{10}$ China Institute of Atomic Energy, Beijing, China\\
$^{11}$ Chungbuk National University, Cheongju, Republic of Korea\\
$^{12}$ Comenius University Bratislava, Faculty of Mathematics, Physics and Informatics, Bratislava, Slovak Republic\\
$^{13}$ COMSATS University Islamabad, Islamabad, Pakistan\\
$^{14}$ Creighton University, Omaha, Nebraska, United States\\
$^{15}$ Department of Physics, Aligarh Muslim University, Aligarh, India\\
$^{16}$ Department of Physics, Pusan National University, Pusan, Republic of Korea\\
$^{17}$ Department of Physics, Sejong University, Seoul, Republic of Korea\\
$^{18}$ Department of Physics, University of California, Berkeley, California, United States\\
$^{19}$ Department of Physics, University of Oslo, Oslo, Norway\\
$^{20}$ Department of Physics and Technology, University of Bergen, Bergen, Norway\\
$^{21}$ Dipartimento di Fisica, Universit\`{a} di Pavia, Pavia, Italy\\
$^{22}$ Dipartimento di Fisica dell'Universit\`{a} and Sezione INFN, Cagliari, Italy\\
$^{23}$ Dipartimento di Fisica dell'Universit\`{a} and Sezione INFN, Trieste, Italy\\
$^{24}$ Dipartimento di Fisica dell'Universit\`{a} and Sezione INFN, Turin, Italy\\
$^{25}$ Dipartimento di Fisica e Astronomia dell'Universit\`{a} and Sezione INFN, Bologna, Italy\\
$^{26}$ Dipartimento di Fisica e Astronomia dell'Universit\`{a} and Sezione INFN, Catania, Italy\\
$^{27}$ Dipartimento di Fisica e Astronomia dell'Universit\`{a} and Sezione INFN, Padova, Italy\\
$^{28}$ Dipartimento di Fisica `E.R.~Caianiello' dell'Universit\`{a} and Gruppo Collegato INFN, Salerno, Italy\\
$^{29}$ Dipartimento DISAT del Politecnico and Sezione INFN, Turin, Italy\\
$^{30}$ Dipartimento di Scienze MIFT, Universit\`{a} di Messina, Messina, Italy\\
$^{31}$ Dipartimento Interateneo di Fisica `M.~Merlin' and Sezione INFN, Bari, Italy\\
$^{32}$ European Organization for Nuclear Research (CERN), Geneva, Switzerland\\
$^{33}$ Faculty of Electrical Engineering, Mechanical Engineering and Naval Architecture, University of Split, Split, Croatia\\
$^{34}$ Faculty of Engineering and Science, Western Norway University of Applied Sciences, Bergen, Norway\\
$^{35}$ Faculty of Nuclear Sciences and Physical Engineering, Czech Technical University in Prague, Prague, Czech Republic\\
$^{36}$ Faculty of Physics, Sofia University, Sofia, Bulgaria\\
$^{37}$ Faculty of Science, P.J.~\v{S}af\'{a}rik University, Ko\v{s}ice, Slovak Republic\\
$^{38}$ Frankfurt Institute for Advanced Studies, Johann Wolfgang Goethe-Universit\"{a}t Frankfurt, Frankfurt, Germany\\
$^{39}$ Fudan University, Shanghai, China\\
$^{40}$ Gangneung-Wonju National University, Gangneung, Republic of Korea\\
$^{41}$ Gauhati University, Department of Physics, Guwahati, India\\
$^{42}$ Helmholtz-Institut f\"{u}r Strahlen- und Kernphysik, Rheinische Friedrich-Wilhelms-Universit\"{a}t Bonn, Bonn, Germany\\
$^{43}$ Helsinki Institute of Physics (HIP), Helsinki, Finland\\
$^{44}$ High Energy Physics Group,  Universidad Aut\'{o}noma de Puebla, Puebla, Mexico\\
$^{45}$ Horia Hulubei National Institute of Physics and Nuclear Engineering, Bucharest, Romania\\
$^{46}$ Indian Institute of Technology Bombay (IIT), Mumbai, India\\
$^{47}$ Indian Institute of Technology Indore, Indore, India\\
$^{48}$ INFN, Laboratori Nazionali di Frascati, Frascati, Italy\\
$^{49}$ INFN, Sezione di Bari, Bari, Italy\\
$^{50}$ INFN, Sezione di Bologna, Bologna, Italy\\
$^{51}$ INFN, Sezione di Cagliari, Cagliari, Italy\\
$^{52}$ INFN, Sezione di Catania, Catania, Italy\\
$^{53}$ INFN, Sezione di Padova, Padova, Italy\\
$^{54}$ INFN, Sezione di Pavia, Pavia, Italy\\
$^{55}$ INFN, Sezione di Torino, Turin, Italy\\
$^{56}$ INFN, Sezione di Trieste, Trieste, Italy\\
$^{57}$ Inha University, Incheon, Republic of Korea\\
$^{58}$ Institute for Gravitational and Subatomic Physics (GRASP), Utrecht University/Nikhef, Utrecht, Netherlands\\
$^{59}$ Institute of Experimental Physics, Slovak Academy of Sciences, Ko\v{s}ice, Slovak Republic\\
$^{60}$ Institute of Physics, Homi Bhabha National Institute, Bhubaneswar, India\\
$^{61}$ Institute of Physics of the Czech Academy of Sciences, Prague, Czech Republic\\
$^{62}$ Institute of Space Science (ISS), Bucharest, Romania\\
$^{63}$ Institut f\"{u}r Kernphysik, Johann Wolfgang Goethe-Universit\"{a}t Frankfurt, Frankfurt, Germany\\
$^{64}$ Instituto de Ciencias Nucleares, Universidad Nacional Aut\'{o}noma de M\'{e}xico, Mexico City, Mexico\\
$^{65}$ Instituto de F\'{i}sica, Universidade Federal do Rio Grande do Sul (UFRGS), Porto Alegre, Brazil\\
$^{66}$ Instituto de F\'{\i}sica, Universidad Nacional Aut\'{o}noma de M\'{e}xico, Mexico City, Mexico\\
$^{67}$ iThemba LABS, National Research Foundation, Somerset West, South Africa\\
$^{68}$ Jeonbuk National University, Jeonju, Republic of Korea\\
$^{69}$ Johann-Wolfgang-Goethe Universit\"{a}t Frankfurt Institut f\"{u}r Informatik, Fachbereich Informatik und Mathematik, Frankfurt, Germany\\
$^{70}$ Korea Institute of Science and Technology Information, Daejeon, Republic of Korea\\
$^{71}$ KTO Karatay University, Konya, Turkey\\
$^{72}$ Laboratoire de Physique Subatomique et de Cosmologie, Universit\'{e} Grenoble-Alpes, CNRS-IN2P3, Grenoble, France\\
$^{73}$ Lawrence Berkeley National Laboratory, Berkeley, California, United States\\
$^{74}$ Lund University Department of Physics, Division of Particle Physics, Lund, Sweden\\
$^{75}$ Nagasaki Institute of Applied Science, Nagasaki, Japan\\
$^{76}$ Nara Women{'}s University (NWU), Nara, Japan\\
$^{77}$ National and Kapodistrian University of Athens, School of Science, Department of Physics , Athens, Greece\\
$^{78}$ National Centre for Nuclear Research, Warsaw, Poland\\
$^{79}$ National Institute of Science Education and Research, Homi Bhabha National Institute, Jatni, India\\
$^{80}$ National Nuclear Research Center, Baku, Azerbaijan\\
$^{81}$ National Research and Innovation Agency - BRIN, Jakarta, Indonesia\\
$^{82}$ Niels Bohr Institute, University of Copenhagen, Copenhagen, Denmark\\
$^{83}$ Nikhef, National institute for subatomic physics, Amsterdam, Netherlands\\
$^{84}$ Nuclear Physics Group, STFC Daresbury Laboratory, Daresbury, United Kingdom\\
$^{85}$ Nuclear Physics Institute of the Czech Academy of Sciences, Husinec-\v{R}e\v{z}, Czech Republic\\
$^{86}$ Oak Ridge National Laboratory, Oak Ridge, Tennessee, United States\\
$^{87}$ Ohio State University, Columbus, Ohio, United States\\
$^{88}$ Physics department, Faculty of science, University of Zagreb, Zagreb, Croatia\\
$^{89}$ Physics Department, Panjab University, Chandigarh, India\\
$^{90}$ Physics Department, University of Jammu, Jammu, India\\
$^{91}$ Physics Department, University of Rajasthan, Jaipur, India\\
$^{92}$ Physics Program and International Institute for Sustainability with Knotted Chiral Meta Matter (SKCM2), Hiroshima University, Hiroshima, Japan\\
$^{93}$ Physikalisches Institut, Eberhard-Karls-Universit\"{a}t T\"{u}bingen, T\"{u}bingen, Germany\\
$^{94}$ Physikalisches Institut, Ruprecht-Karls-Universit\"{a}t Heidelberg, Heidelberg, Germany\\
$^{95}$ Physik Department, Technische Universit\"{a}t M\"{u}nchen, Munich, Germany\\
$^{96}$ Politecnico di Bari and Sezione INFN, Bari, Italy\\
$^{97}$ Research Division and ExtreMe Matter Institute EMMI, GSI Helmholtzzentrum f\"ur Schwerionenforschung GmbH, Darmstadt, Germany\\
$^{98}$ Saga University, Saga, Japan\\
$^{99}$ Saha Institute of Nuclear Physics, Homi Bhabha National Institute, Kolkata, India\\
$^{100}$ School of Physics and Astronomy, University of Birmingham, Birmingham, United Kingdom\\
$^{101}$ Secci\'{o}n F\'{\i}sica, Departamento de Ciencias, Pontificia Universidad Cat\'{o}lica del Per\'{u}, Lima, Peru\\
$^{102}$ Stefan Meyer Institut f\"{u}r Subatomare Physik (SMI), Vienna, Austria\\
$^{103}$ SUBATECH, IMT Atlantique, Nantes Universit\'{e}, CNRS-IN2P3, Nantes, France\\
$^{104}$ Suranaree University of Technology, Nakhon Ratchasima, Thailand\\
$^{105}$ Technical University of Ko\v{s}ice, Ko\v{s}ice, Slovak Republic\\
$^{106}$ The Henryk Niewodniczanski Institute of Nuclear Physics, Polish Academy of Sciences, Cracow, Poland\\
$^{107}$ The University of Texas at Austin, Austin, Texas, United States\\
$^{108}$ Universidad Aut\'{o}noma de Sinaloa, Culiac\'{a}n, Mexico\\
$^{109}$ Universidade de S\~{a}o Paulo (USP), S\~{a}o Paulo, Brazil\\
$^{110}$ Universidade Estadual de Campinas (UNICAMP), Campinas, Brazil\\
$^{111}$ Universidade Federal do ABC, Santo Andre, Brazil\\
$^{112}$ University of Cape Town, Cape Town, South Africa\\
$^{113}$ University of Houston, Houston, Texas, United States\\
$^{114}$ University of Jyv\"{a}skyl\"{a}, Jyv\"{a}skyl\"{a}, Finland\\
$^{115}$ University of Kansas, Lawrence, Kansas, United States\\
$^{116}$ University of Liverpool, Liverpool, United Kingdom\\
$^{117}$ University of Science and Technology of China, Hefei, China\\
$^{118}$ University of South-Eastern Norway, Kongsberg, Norway\\
$^{119}$ University of Tennessee, Knoxville, Tennessee, United States\\
$^{120}$ University of the Witwatersrand, Johannesburg, South Africa\\
$^{121}$ University of Tokyo, Tokyo, Japan\\
$^{122}$ University of Tsukuba, Tsukuba, Japan\\
$^{123}$ University Politehnica of Bucharest, Bucharest, Romania\\
$^{124}$ Universit\'{e} Clermont Auvergne, CNRS/IN2P3, LPC, Clermont-Ferrand, France\\
$^{125}$ Universit\'{e} de Lyon, CNRS/IN2P3, Institut de Physique des 2 Infinis de Lyon, Lyon, France\\
$^{126}$ Universit\'{e} de Strasbourg, CNRS, IPHC UMR 7178, F-67000 Strasbourg, France, Strasbourg, France\\
$^{127}$ Universit\'{e} Paris-Saclay, Centre d'Etudes de Saclay (CEA), IRFU, D\'{e}partment de Physique Nucl\'{e}aire (DPhN), Saclay, France\\
$^{128}$ Universit\'{e}  Paris-Saclay, CNRS/IN2P3, IJCLab, Orsay, France\\
$^{129}$ Universit\`{a} degli Studi di Foggia, Foggia, Italy\\
$^{130}$ Universit\`{a} del Piemonte Orientale, Vercelli, Italy\\
$^{131}$ Universit\`{a} di Brescia, Brescia, Italy\\
$^{132}$ Variable Energy Cyclotron Centre, Homi Bhabha National Institute, Kolkata, India\\
$^{133}$ Warsaw University of Technology, Warsaw, Poland\\
$^{134}$ Wayne State University, Detroit, Michigan, United States\\
$^{135}$ Westf\"{a}lische Wilhelms-Universit\"{a}t M\"{u}nster, Institut f\"{u}r Kernphysik, M\"{u}nster, Germany\\
$^{136}$ Wigner Research Centre for Physics, Budapest, Hungary\\
$^{137}$ Yale University, New Haven, Connecticut, United States\\
$^{138}$ Yonsei University, Seoul, Republic of Korea\\
$^{139}$  Zentrum  f\"{u}r Technologie und Transfer (ZTT), Worms, Germany\\
$^{140}$ Affiliated with an institute covered by a cooperation agreement with CERN\\
$^{141}$ Affiliated with an international laboratory covered by a cooperation agreement with CERN.\\

\end{flushleft} 

\end{document}